%% file: 2011-cw+kondo.tex
\documentclass[prb,twocolumn,floatfix,footinbib,superscriptaddress,showpacs, showkeys,notitlepage]{revtex4-1}
\usepackage{graphicx}
\usepackage{amssymb}
\usepackage[utf8]{inputenc}
\usepackage[english]{babel}
\usepackage[usenames,dvipsnames]{color}
\usepackage{longtable}
\usepackage{array}
\usepackage{multirow}
\usepackage{hyperref}
\newcommand{\wn}{\ensuremath{\omega_n}}

\newcommand{\e}{\epsilon}
\newcommand{\up}{\uparrow}
\newcommand{\down}{\downarrow}
\newcommand{\updown}{\up\down}
\newcommand{\gs}{\Psi}

\newcommand{\ket}[1]{|#1\rangle}
\newcommand{\biket}[2]{\ket{#1;#2}}
\newcommand{\triket}[3]{\ket{#1;#2;#3}}
\newcommand{\melem}[3]{\langle#1|#2|#3\rangle}

\usepackage{amsmath}

\DeclareMathOperator*{\sgn}{sgn}

\begin{document}
\title{Role of rotational symmetry in the magnetism of a multiorbital model}
\author{A.~E.~Antipov}\email{antipov@pks.mpg.de}
\affiliation{Max Planck Institute for Chemical Physics of Solids, N\"othnitzer Stra\ss{}e 40, 01187 Dresden, Germany}
\affiliation{Max Planck Institute for the Physics of Complex Systems, N\"othnitzer Stra\ss{}e 38, 01187 Dresden, Germany}
\author{I.~S.~Krivenko}
\affiliation{Centre de Physique Th\'{e}orique, Ecole Polytechnique, CNRS, 91128 Palaiseau Cedex, France}
\author{V.~I.~Anisimov}
\affiliation{Institute of Metal Physics, Sofia Kovalevskaya Street 18, 620219 Ekaterinburg GSP-170, Russia}
\affiliation{Theoretical Physics and Applied Mathematics Department, Urals Federal University, Mira Street 19,  620002
Ekaterinburg, Russia}
\author{A.~I.~Lichtenstein}
\affiliation{Institut f\"u{}r Theoretische Physik, Universitat Hamburg, Jungiusstra\ss{}e 9, D-20355 Hamburg, Germany}
\author{A.~N.~Rubtsov}
\affiliation{Department of Physics, Moscow State University, 119992 Moscow, Russia}

\date{\today}

\begin{abstract}
Effect of rotationally-invariant Hund's rule coupling on a magnetism of multiorbital Hubbard models is studied within a dynamical mean field theory framework.
Comparison of static magnetic susceptibilities and local densities of states of two- and three-orbital models of a complete rotationally invariant Coulomb interaction and a ``density-density'' Hartree type interaction shows the different role of spin-flip interactions for different band fillings.
In the particle-hole symmetric case the Mott-Hubbard physics dominates due to the strong effective Coulomb interaction, while for the multiple electronic configurations away from half-filling (two electrons in the three band model) the formation of local magnetic moments due to Hund's exchange interaction becomes the most significant effect for itinerant magnetic systems.
A shift of the temperature of magnetic ordering due to the rotationally-invariant Hund's rule coupling is found to be the largest in a three-orbital model with a two-electron occupancy where the single particle spectrum is metallic and is not sensitive to different forms of the Coulomb vertex.
A larger enhancement of the effective mass in a model with a rotationally-invariant interaction is discussed.
In the half-filled case we find a drastic change in the density of states close to the Mott transition which is related to the spin-flip Kondo fluctuations in a degenerate orbital case, while the corresponding shift of the magnetic transition temperature is relatively small.
It is shown that a change in the ground state degeneracy due to a different symmetry of the Coulomb interaction in the density-density model leads to a breakdown of the quasiparticle peak at the Fermi level in the proximity of a Mott transition on the metallic side.
We discuss the relevance of rotationally-invariant Hund's interaction in the  transition metal magnetism.
\end{abstract}


\maketitle

\section{Introduction}

In many materials with strong electronic correlations spin and orbital fluctuations play an important role\cite{ImadaFujimori:1998}. Experimental spectroscopic data\cite{MarelSawatsky:1988} and analysis of a simple model\cite{Norman:1995} show that the local Coulomb interaction of valence electrons is sufficiently screened, while the higher-order Slater integrals are almost unaffected by a metallic screening. 
This gives rise to  important effects, associated with the Hund's exchange coupling $J$ in strongly-correlated magnetic compounds. In the case of relatively large onsite Coulomb repulsion in transition metal compounds a ferromagnetic state can be stabilized by orbital ordering of the orbitals in the valence shell\cite{KugelKhomskii:1982, Fazekas:2000}.
A variation of an effective hybridization between different orbitals can lead to an orbital-selective Mott transition \cite{Anisimov:2002, KogaKawakami:2004, Liebsch:2005, CostiLiebsch:2007} which was observed experimentally in  La$_{n+1}$Ni$_{n}$O$_{3n+1}$\cite{SreedharMcElfresh:1994}.
In the strongly correlated metals away from the particle-hole symmetry the effective Coulomb interaction is well screened and the relevant electron interaction is related to a Hund's rule coupling with complicated multiplet fluctuations \cite{MediciMravlje:2011}. 
The physics of such Hund's metals is dominated by the spin-flip type of fluctuations which lead to a metallic state with correlated local-moments and a reduced spectral weight \cite{MediciMravlje:2011, Mravlje:2011}.

The effective way of making a non-perturbative description of strongly correlated electronic systems is a self-consistent mapping of a complex lattice problem to a single impurity with an energy-dependent external bath.
In this dynamical mean field theory (DMFT) \cite{GeorgesKotliar:1996} the resulting Anderson impurity problem is solved numerically. 
Until recently  numerical solvers for this problem could treat the Hartree-type density-density Coulomb interaction, while an inclusion of the off-diagonal terms led to an exponential increase of the computational time\cite{PruschkeBulla:2005, SakaiArita:2006}. 
A significant progress has been achieved with the appearance of the continuous-time quantum Monte-Carlo algorithms \cite{GullMillis:2011}. 
An application of the algorithm to a single Anderson impurity problem with a $3d$ valence shell (Co atom on a metallic surface) and taking into account a full multiorbital character of the Coulomb interaction results in a shift of the Kondo peak in the density of states and a reduction of the Kondo temperature \cite{GorelovWehling:2009, SurerTroyer:2012, NevidomskyyColeman:2009}. 

A realistic description of the strongly-correlated electron materials such as transition metal systems with partially filled $3d$ bands is often impossible without  taking into account multiplet effects in the valence shell. 
Indeed previous LDA+DMFT calculations for the density-density type interaction overestimate the Curie temperature of  iron\cite{LichtensteinKatsnelson:2001} and various manganites\cite{HeldVollhardt:2000} by a factor of $1.5-2$. 
This inconsistency can be associated either with the absence of non-local correlations in the dynamical mean-field  scheme \cite{FlorensGeorges:2002} or with the lack of spin-flip and pair-hopping terms which are present in the realistic multiorbital Hubbard-like model \cite{Rozenberg:1997,KotliarSavrasov:2006}. 
In order to separate these contributions we study the Hubbard model in the limit of infinite dimensions where the local dynamical mean field theory becomes exact \cite{MetznerVollhardt:1989,GeorgesKotliar:1996}.

In recent years an extensive study of the two- and three-orbital Hubbard models in infinite dimensions with Numerical Renormalization group (NRG) \cite{PruschkeBulla:2005, BullaCosti:2008, PetersPruschke:2010, PetersKawakami:2011}, self-energy functional approach \cite{InabaKoga:2007}, equation of motion method \cite{FengOppeneer:2012}, highly optimized Hirsch-Fye\cite{SakaiArita:2006}, exact diagonalization \cite{LiebschIshida:2012} and continuous time Quantum Monte Carlo (CT-QMC) \cite{WernerMillis:2006, WernerMillis:2007, WernerGull:2008, WernerGullMillis:2009, ChanWerner:2009} solvers has been performed and the phase diagrams of the models are established. 
Insulating phases occur at integer electronic fillings. 
In the half-filled case the antiferromagnetic behavior is observed for any $U$ larger than the critical value of the metal-insulator phase transition, which is of the order of bandwidth. 
Away from particle-symmetric regime of the model a ferromagnetism coupled with the orbital order of the electrons in the lattice can be stabilized at large values of $U$, otherwise a paramagnet or in the case of filling with $2$ electrons in a $3$-orbital shell the frozen local moment\cite{WernerGull:2008} phase is observed. 
An application of the model to various transition metal oxides with a large Hund's coupling reveal a strong enhancement of the effective mass depending on the valence band electronic fillings\cite{MediciMravlje:2011}.  

Despite of a general understanding of the low-temperature phase diagram of the multiorbital Hubbard model, a careful comparison with the simplified density-density Coulomb interaction model, which is generally used in applications to real materials, needs to be performed. 
One of the experimentally accessible observable quantities is the critical temperature of magnetic phase transitions. 
In our recent letter [\onlinecite{AAA:2011}] the static magnetic susceptibility of the half-filled two-orbital paramagnetic Hubbard model was obtained using the continuous-time interaction-expansion quantum Monte-Carlo solver\cite{RubtsovSavkin:2005}.

In this paper we discuss the dependence of the magnetic phase transition on the electronic filling in two- and three-orbital Hubbard models for the general rotationally invariant and density-density Coulomb interactions. The interaction expansion variant of the continuous-time quantum Monte-Carlo solver \cite{RubtsovSavkin:2005,GullMillis:2011} is used for the calculations. 
The paper is organized as follows. In section II the general multiorbital rotationally invariant Hubbard-like model is introduced.
In section III the results for the static magnetic susceptibilities for two- and three- orbital models on an infinite-dimensional Bethe lattice for different electronic fillings are presented. 
We discuss the importance of spin-flip Hund's rule effects in the half-filled model in section IV and away from half-filling in section V. 
In appendix \ref{DF} we show that the reduced degeneracy of the ground state of the ``density-density'' model leads to the disappearance of a quasiparticle peak at the Fermi level. This behavior is effectively traced in a hybridization expansion in auxiliary dual fermions, which uses vertex functions of the ``atomic'' problem as a building block for the perturbation theory \cite{HafermannLi:2009, 
KrivenkoRubtsov:2010}.

\section{Model}
The multiorbital extension of the Hubbard model can be written in the following form:
\begin{multline}
\hat H = -t\sum_{\langle ij\rangle\alpha\sigma}(\hat
c^+_{i \alpha\sigma}\hat c_{j \alpha\sigma} + \hat c^+_{j \alpha\sigma}\hat c_{i \alpha\sigma}) - \\ - \mu\sum_{i \alpha\sigma}\hat n_{i \alpha\sigma}
+ \sum_{i} \hat H^{\text{int}}_{i},
\end{multline} 
where $t$ denotes the (orbital independent) hopping between adjacent sites $i$ and $j$ of the lattice and sets the energetic scale of the model, $\alpha$ and $\sigma$ are orbital and spin indices, $\mu$ is the chemical potential and $\hat H^{\text{int}}$ is the local onsite Coulomb interaction. Without symmetry breaking the Hamiltonian of the onsite Coulomb interaction obeys the rotational invariance and the conservation of spin and can be expressed in terms of the squared total spin $\hat S^2$, squared angular momentum $\hat L^2$ and number of particles $\hat N$ operators. For the shell with conserved $l=1$ angular momentum and $s=1/2$ spin of each electron it reads\cite{Racah:1942}
\begin{multline}\label{RotInvU}
    \hat H_{int} =\\= \left(4J-\frac{U}{2}\right)\hat N + \left(U - 3J\right)\frac{\hat N^2}{2}
        -J\left[2\hat S^2 + \frac{\hat L^2}{2}\right].
\end{multline}
Being expressed in the basis of Fock states the Hamiltonian (\ref{RotInvU}) takes the form introduced in Ref. [\onlinecite{Oles:1983}] and can be split into a part which corresponds to a density-density type of the interaction and the remaining off-diagonal terms, which include spin-flip and pair-hopping terms:
\begin{flalign}\label{Umatrix}
&\hat H^{\text{int}} = \hat H^{\text{nn}} + \hat H^{\text{sf}} &
\\ \label{DensityDensityMatrix}
&\hat H^{\text{nn}} = \frac{U}{2}\sum_{\alpha \sigma \neq \sigma'} n_{\alpha\sigma}n_{\alpha\bar\sigma} +
		\frac{U-2J}{2} \sum_{\alpha\neq \alpha',\sigma} n_{\alpha\sigma} n_{\alpha'\bar\sigma} + & \\ \notag
    &+	\frac{U-3J}{2} \sum_{\alpha\neq \alpha',\sigma} n_{\alpha\sigma} n_{\alpha'\sigma}& \\ \notag
&\hat H^{\text{sf}}=- \frac{J}{2}\sum_{\alpha\neq \alpha',\sigma} 
	(c^\dagger_{\alpha \sigma}c^\dagger_{\alpha'\bar\sigma}c_{\alpha'\sigma}c_{\alpha\bar\sigma} +	c^\dagger_{\alpha'\sigma}c^\dagger_{\alpha'\bar\sigma}c_{\alpha\sigma}c_{\alpha\bar\sigma}).&
\end{flalign}
Here $\bar\sigma$ denotes the opposite of $\sigma$. The model (\ref{Umatrix}) keeps its degeneracy in a cubically symmetric environment and as such (with an introduction of parameter $U'$, here equal to $U-2J$) is used for a description of a three-orbital $t_{2g}$ and also two-orbital $e_g$ electronic shells \cite{MizokawaFujimori:1995}, obtained for crystal field splitting of the $d$-shell. Note that equation (\ref{RotInvU}) is not applicable for the two-orbital case.

\input{tables.tex}

Several remarks should be done about the single-atomic limit of the model. At $t=0$ rotationally invariant and ``density-density'' models have different degeneracy of the ground states in various electronic occupation sectors. The spectrum of the model is shown in tables \ref{2bandTable} and \ref{3bandTable} for two- and three- orbital complexity respectively. The tables are analogous to the ones in Ref. \onlinecite{WernerMillis:2006} and supplementary material of Ref. \onlinecite{MediciMravlje:2011}.  These are exactly the pair-hopping and spin-flip terms which entangle the Fock states in (\ref{Umatrix}) and make the spin triplet a ground state in a half-filled rotationally invariant model. As it was pointed out by Pruschke and Bulla\cite{PruschkeBulla:2005} the change in the character of the ground states has a profound impact on the excitation spectra. Kondo-type spin-flip processes are clearly possible in the atom with a multiplet classification of multielectron levels, where $S_z$ can be changed to $S_z\pm 1$, preserving the atomic energy. Hund's rule guarantees that the ground state in a rotationally invariant model has the largest possible $\langle S^2\
\rangle$, so that
the processes $S_z\to S_z\pm 1$ are indeed allowed and the contribution to the central quasiparticle peak of Kondo scattering is present.

In the ``density-density'' case the ground state has a lower degree of degeneracy and consists only of Fock states with all spins aligned in one direction which corresponds to the Ising-type interaction. The spin projections of different ground states differ by $\Delta S_z=1$ for the 2-band model and by $\Delta S_z=3/2$ for the 3-band model, so no scattering process in the impurity can induce a gapless Kondo excitation.
This behavior is observed in the excitation spectra of the lattice models in the vicinity of a Mott transition and is discussed in section IV and its analytic treatment in this case is presented in Appendix \ref{DF}.

\begin{figure}[t!]
\includegraphics[width=\columnwidth]{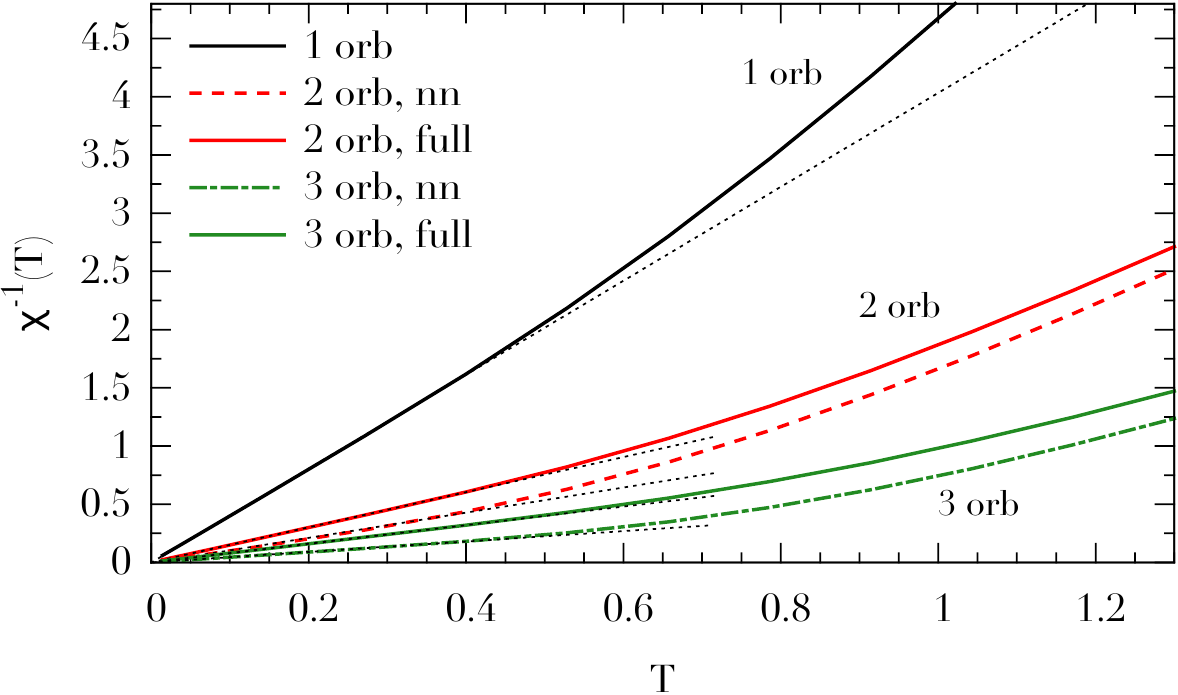}
\caption{(Color online) The temperature dependence of the inverse magnetic susceptibility of a single unhybridized atomic orbital, defined by full rotationally invariant Hamiltonian (\ref{Umatrix}) (solid lines) and (\ref{DensityDensityMatrix}) (dashed lines) in the particle-hole symmetric regime at $U=3.6, J=1$ for orbital dimensionalities $1,2,3$.}
\label{CwED}
\end{figure}

The change in the degree of degeneracy of the ground state of the ``atomic'' solution of the model has a clear consequence on the Curie-law dependence of inverse magnetic susceptibility on temperature, shown in Fig. \ref{CwED}. 
In the low-temperature regime $T \ll J < U$ the Curie constant is obtained as a response of a single $S=1$ or $3/2$ spin for two- and three-orbital models respectively. The density-density type Hamiltonian possesses only states with $|S_z| = S$ and thus has larger Curie constant and magnetic susceptibility. Indeed at $T<J$, $C_{N=2}^{\text{full}} = 2/3$, $C_{N=2}^{\text{nn}}=1$, $C_{N=3}^{\text{full}}=5/4$, $C_{N=3}^{\text{nn}}=9/5$. In the high-temperature regime weights of all states become the same and the Curie constant is defined solely by the dimensionality of the Hilbert space and has no difference between density-density and rotationally invariant models.

\section{Static magnetic susceptibility calculation}
We proceed now with the lattice calculations. In the limit of infinite dimensions the linked-cluster interaction expansion\cite{Metzner:1991} of the Hubbard model is reduced to the analogous of the Anderson impurity model\cite{GeorgesKotliar:1996}, with the effective hybridization function $\Delta_{\alpha\sigma}(i\wn)$ equal to 
\begin{equation}\label{Bethe-SC}
\Delta_{\alpha\sigma}(i\wn) = t^2 g_{\alpha\bar\sigma}(i\wn),
\end{equation}
where the spin-polarization can be imposed with $\bar\sigma$ index to stabilize the anti-ferromagnetic ordering. The action for the lattice problem then can be replaced with the following:
\begin{multline}\label{SIAMcw}
S^{\text{imp}} = -\sum_{\alpha\sigma}\sum_{\wn}(i\wn + \mu -\Delta_{\alpha\sigma}(i\wn))c^*_{\wn \alpha\sigma}c_{\wn \alpha\sigma} +\\+ \int\limits_0^\beta d\tau H^{\text{int}}[c,c^*] - 2\int d\tau h \hat S_z (\tau),
\end{multline}
where $\wn$ are Matsubara frequencies, $\hat S_z = \frac{1}{2}\sum_{\alpha}  (\hat n_{\alpha \uparrow} - \hat n_{\alpha \downarrow})$ is the total spin projection along the $z$ axis and electronic magnetic moment is taken as unity. In order to obtain the static magnetic susceptibility a small external magnetic field $h$ is applied along $z$ axis. The magnetic susceptibility then reads:
\begin{equation}\label{CWdef}
\chi_q = 2\frac{d{\langle \hat S_z \rangle}}{d h_q}\Bigr|_{h\rightarrow 0},
\end{equation}
where $q=0,X$ are generalized Brilloin zone points corresponding to ferromagnetic $(0)$ or antiferromagnetic  $(X)$ staggered magnetic field.
Finally, the  critical (Curie/Neel) temperature is obtained by a linear interpolation of the Curie-Weiss law:
\begin{equation}\label{CW}
\chi = \frac{C}{T-T_c},
\end{equation}
where $C$ is the Curie constant. Since at the limit of infinite dimensions the lattice effects are included at the dynamical mean-field level, the Weiss temperature obtained in the aforementioned way is indeed the magnetic ordering temperature\cite{ByczukVollhardt:2002}. Note that it is also positive for the antiferromagnetic ordering as a response of bipartite  (\ref{Bethe-SC}) lattice on the staggered field (\ref{CWdef}) applied with $\vec q=X$. This corresponds to the $\chi(\vec q = X)$ susceptibility in finite-dimensional lattices.

\begin{figure}[t]
\includegraphics[width=\columnwidth]{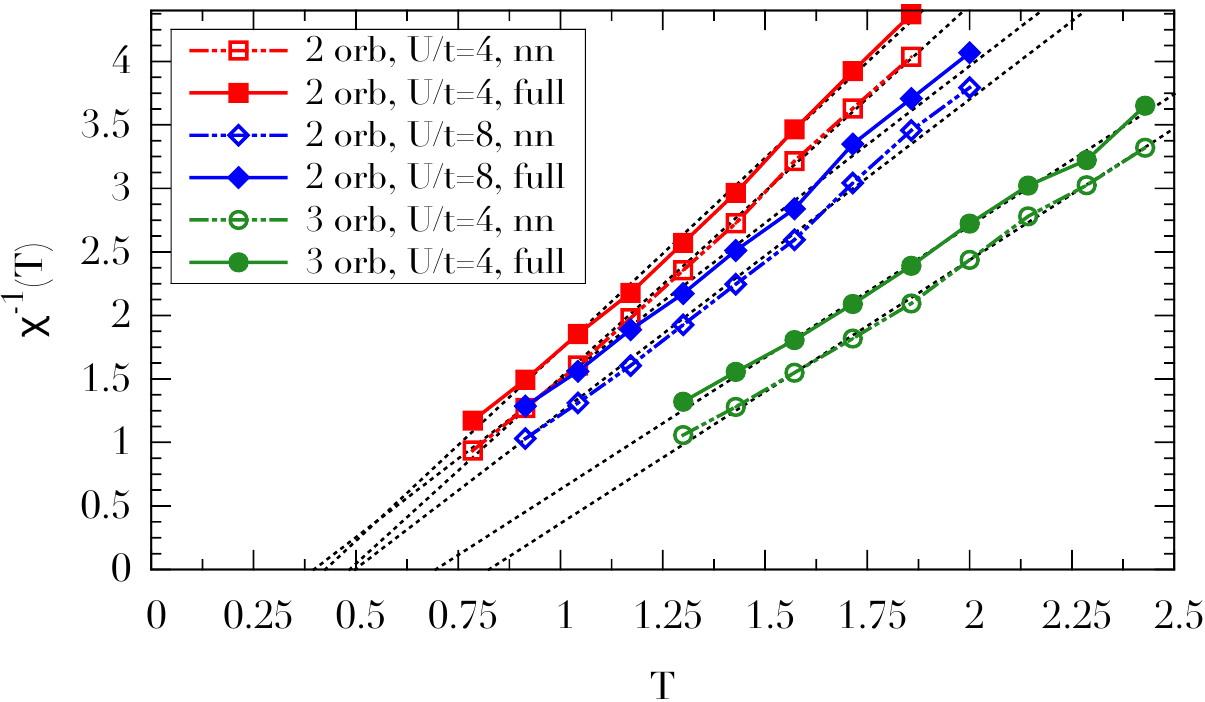}
\caption{(Color online) Curie-Weiss dependence of the inverse magnetic susceptibility $\chi^{-1}(T)$ over temperature of the half-filled two- and three-orbital Hubbard model at $U/t=4, J=1.2t$ and $U=8t, J=1.2t$ (for the two-orbital case only). The curves for the rotationally invariant interaction are named "full" and plotted with straight lines and filled points, while the curves which correspond to the density-density type of interaction are named "nn" and plotted with dashed lines and empty points. The dotted lines are linear fits of the inverse susceptibilities in the paramagnetic regime. The Neel temperature is defined as the crossing point of interpolation lines with the $\chi^{-1}(T)=0$ axis. The temperatures obtained are $T^{\text{full}}_{N=2} = 0.42$ in the rotationally invariant model and $T^{\text{nn}}_{N=2} =0.49$ in the density-density variant of the two-orbital model. In the three-orbital case the corresponding values are $T^{\text{full}}_{N=3} = 0.7$ and $T^{\text{nn}}_{N=3} =0.83$.}
\label{cw_hfilled}
\end{figure}

The results for the inverse of the static magnetic susceptibility as a function of temperature for the models with full rotational invariance and density-density type of the interaction in the particle-hole symmetric case are plotted on Fig. \ref{cw_hfilled}. First, it is seen that exactly as in the case of an "atomic solution" on Fig. \ref{CwED} an increase of the total number of orbitals leads to an enhancement of the static magnetic susceptibility in the half-filled case, yet in the lattice it is also followed by a corresponding increase of the Neel temperature. The ratio between $T_c$ in half-filled rotationally-invariant two and three orbital models is equal to $1.7$ which is a bit less than the $15/8$ ratio of  the mean field critical temperatures in the Heisenberg model, for which the magnetic ordering temperature in the mean field approximation is defined as $T_c=\mathcal{J}zS(S+1)/3$ with the magnetic exchange constant $\mathcal{J}$ and $z$ nearest neighbors in the lattice. 
The difference between these values is associated with the charge fluctuations absent in the Heisenberg treatment. 
Comparing curves for the two-orbital model at $U/t=8$ and $U/t=4$ with the same value of $J=1.2t$ reveals that an increase of Hubbard $U$ and thus a reduction of the effect of charge fluctuations leads to a small change in the Curie constant and does not affect the Neel temperature in this regime.

The important consequence of taking into account spin-flip and pair-hopping terms is that it leads to a decrease of a magnetic ordering temperature. It is clear that the spin-flip terms generally help to rotate large local magnetic moments and reduce the critical temperature.  This behavior has a prototype in a comparison of critical temperatures of Ising and Heisenberg models. In the half-filled case on Fig. \ref{cw_hfilled} the magnitude of this effect is approximately $20\%$ both for two- and three-orbital models. 

\begin{figure}[t]
\includegraphics[width=\columnwidth]{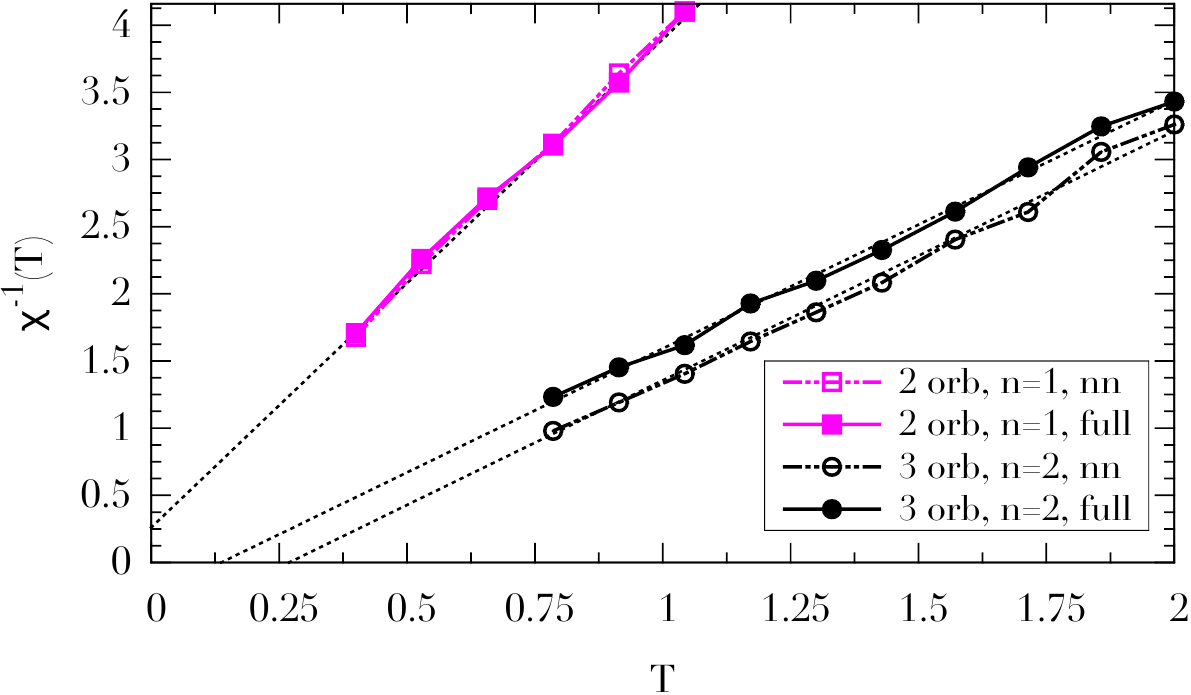}
\caption{(Color online) Curie-Weiss dependence of the inverse magnetic susceptibility $\chi^{-1}(T)$ over temperature of the two-orbital model at the single-electron filling and the three-orbital model at two-electron filling. The curves are obtained at $U=8, J=1.2t$. The notations are the same as in Fig. \ref{cw_hfilled}. The obtained magnetic transition temperatures in the three-orbital model at one-third filling differ by a factor of $2$ and have values $T^{\text{full}}_{N=3; n=2} = 0.14$ in the rotationally invariant model and $T^{\text{nn}}_{N=3; n=2} = 0.27$ in the ``density-density'' model. }
\label{cwFM}
\end{figure}

The curves for the inverse of the static magnetic susceptibility as a function of temperature away from half-filling are plotted on Fig. \ref{cwFM}. 
The case of the one-third filling (with $n=2$ electrons) in the three orbital model is of particular interest. Previously it has been shown that a state with a frozen moment is realized at low temperatures at moderate values of $U$ \cite{WernerGull:2008}, whereas an increase of $U$ leads to the ferromagnetic spin ordering\cite{ChanWerner:2009}. We thus impose a self-consistency condition (\ref{Bethe-SC}) without spin inversion in this case \footnote{Note that the self-consistency itself only affects the stability of the solution in a proximity to the phase transition, while the high-temperature paramagnetic behavior remains the same}. It is first seen, that the static uniform magnetic susceptibility is larger than in the half-filled case. An explanation for this fact is that the degree of degeneracy of the ground state and thus the magnetic moment of the shell with two electrons is larger than of the half-filled one by a combinatorial factor (see Table \ref{3bandTable}). Larger magnetic susceptibility in the one-third filled three-orbital shell leads to an enhanced ratio between Curie temperatures of the density-density and rotationally invariant Coulomb interaction models which in the described case reaches a factor of two. This enhancement is related to the fact that correlation effects are almost completely dominated by the Hund's parameter J, while the effective Coulomb interaction is sufficiently reduced. The difference between the half-filled and one-third filled cases is covered in greater detail in Discussion.

A large doping of the system which corresponds to $n=1$ average electrons number per site in the two orbital model leads to the obvious result. It is seen on Fig. \ref{cwFM} that in this case no spin-flip or pair-hopping processes which involve two electrons can occur, and thus there is no difference between rotationally invariant and density-density models. The fact that the line of the inverse magnetic susceptibility crosses $0$ at a negative value corresponds to the fact that the model in the described regime stays paramagnetic at zero temperature\cite{PetersKawakami:2011}.

\section{Density of states}

\begin{figure}[t]
\flushleft{(a)}\\
\includegraphics[width=0.96\columnwidth]{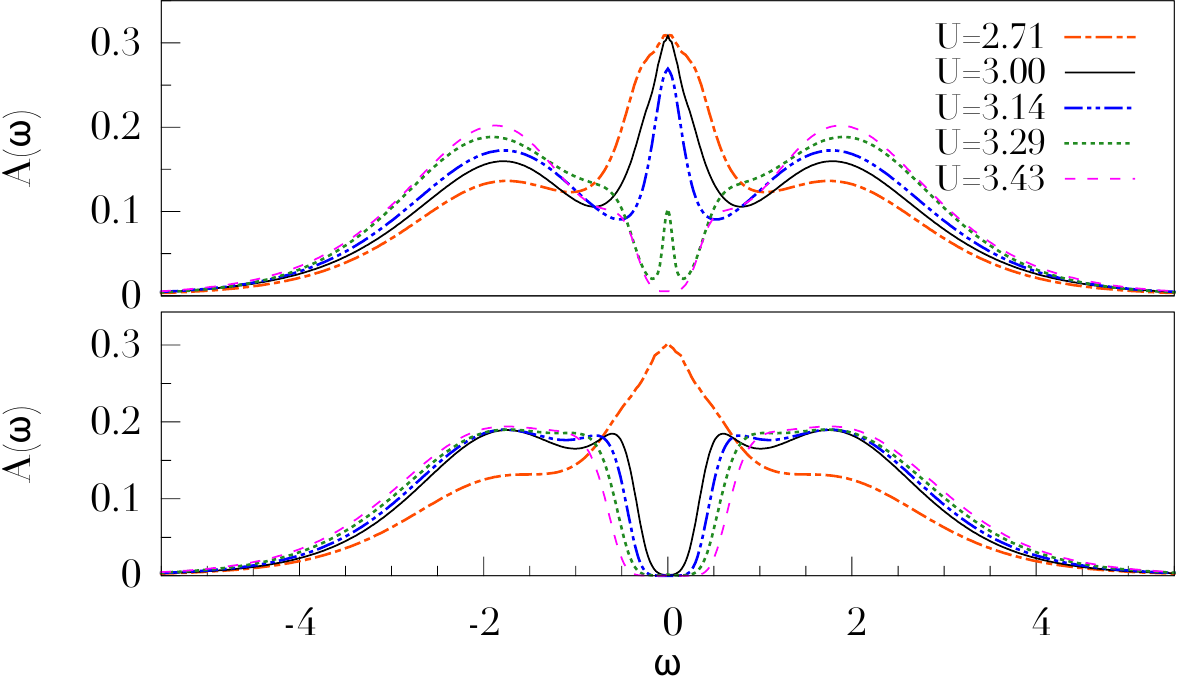} \\\vspace*{-2.1em}
\flushleft{(b) \\ 
\includegraphics[width=0.96\columnwidth]{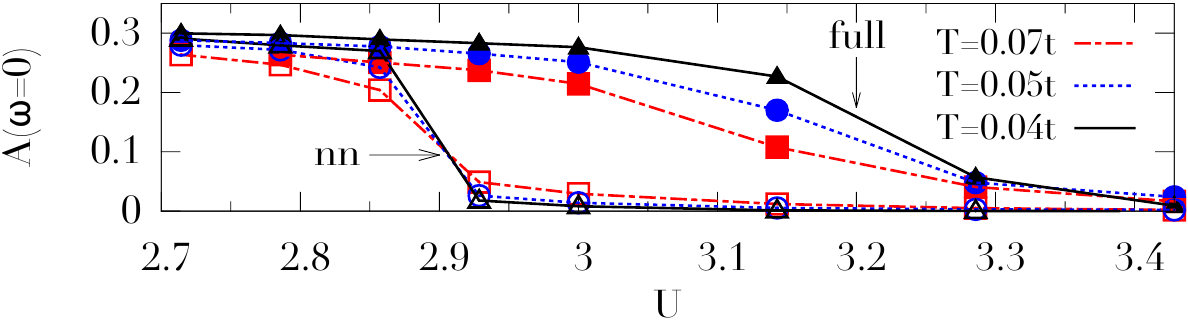} } 
\caption{(a) Family of densities of states of the two-orbital Hubbard model for various values of $U$ in the vicinity of the metal-insulator transition, $J=U/4, T=0.04t$. All curves are produced by using the analytic continuation by the maximum entropy method\cite{Sandvik:1998}. The upper part of the graph contains results for the rotationally-invariant Coulomb interaction (\ref{Umatrix}), the lower --- for its density-density part (\ref{DensityDensityMatrix}). (b) Density of states at the Fermi level $A(0)$ for rotationally invariant and density-density models as a function of $U$, $J=U/4$ at temperatures $0.04t, 0.05t, 0.07t$. The points were obtained from the extrapolation of Matsubara Green's function to the zero value and showed no difference from the value $\beta G(\tau=\beta/2) = A(\omega = 0)|_{T\rightarrow 0}$.}\label{MultipleDOS2}
\end{figure}

In this section we study the low-temperature behavior of rotationally invariant and density-density models and probe the Mott transition by changing the onsite interaction parameter $U$ in two and three-orbital models at various fixed temperatures at half-filling.
In order for the Mott transition not to get hidden by the antiferromagnetism the paramagnetic model is studied and the symmetry against spin inversion in the effective impurity model is enforced.
The external magnetic field is kept at zero value.

Densities of states for the two-orbital model in the vicinity of a Mott transition obtained by analytic continuation\cite{Sandvik:1998} for various values of $U$ are plotted on Fig. \ref{MultipleDOS2}a.
The parameters are chosen to be the same as in Fig. $7$ of Ref. \onlinecite{PruschkeBulla:2005} and $J$ is fixed at $U/4$, yet the finite-temperature behavior at $T=0.04t$ is described here. 
The curves for the rotationally invariant interaction (\ref{Umatrix}) are  reminiscent of the single-band Hubbard model and at $U<U_c \approx 3.3t $ show a quasiparticle peak at the Fermi-level, coupled with shoulders from the Coulomb interaction at $\pm U/2$, whereas at $U>U_c$ the charge gap in the density of states is opened. The finite temperature effects lead to the lack of ``pinning'' of the central quasiparticle peak and thus its height is not fixed for different values of $U$ in the metallic side. This effect is clearly seen on Fig. \ref{MultipleDOS2}b, where the value of density of states at Fermi level is plotted as a function of $U$ at different temperatures. Here a decrease of the temperature leads to ``straightening'' of the line of $A(\omega = 0)$ values in the metallic part and so one expects the
``uncorrelated'' value of DOS to be restored at zero temperature.

\begin{figure}[t]
\flushleft
\includegraphics[width=1.0\columnwidth]{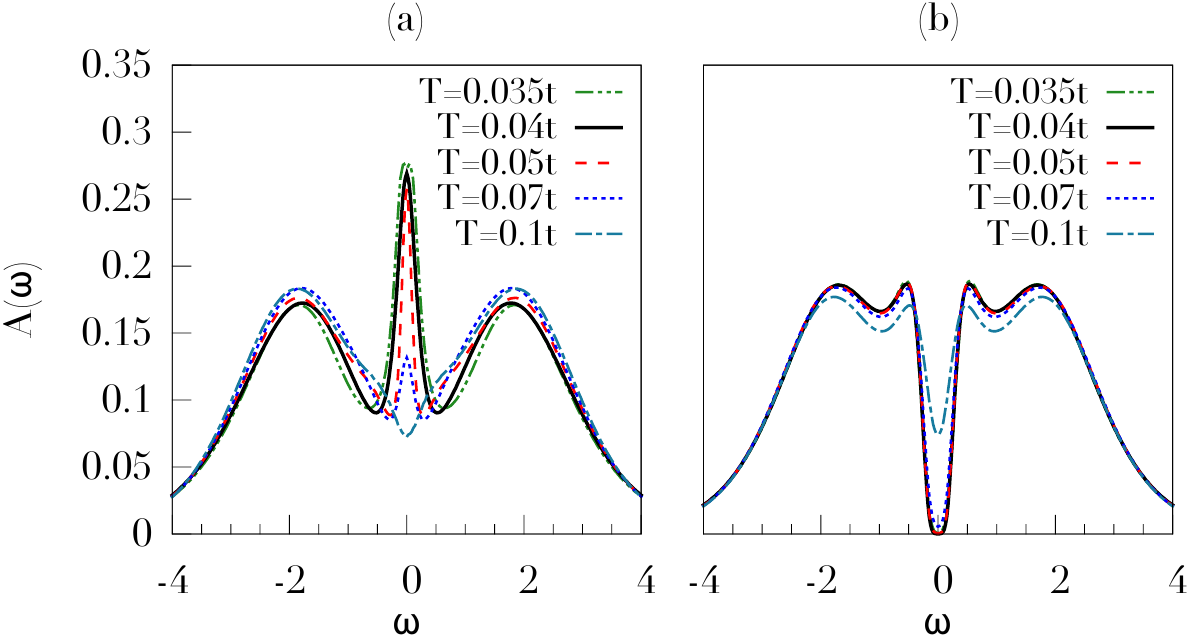}
\caption{Family of densities of states of the two-orbital Hubbard model for various values of temperature for the models with the (a) rotationally-invariant Coulomb interaction at $U=3.14t, J=U/4$, (b) density-density interaction at $U=2.9t, J=U/4$. }\label{DOS2_T}
\end{figure}

A comparison of the rotationally invariant and ``density-density'' curves at Fig. \ref{MultipleDOS2}a and \ref{MultipleDOS2}b reveals the same effect previously observed at zero temperature \cite{PruschkeBulla:2005}, namely that taking into account only ``density-density'' part of the interaction leads to a breakdown of the central quasiparticle peak at Fermi level and thus to a sufficient reduction of the critical $U$ to $\approx 2.9t$. In order to explore the difference between models in greater detail we fix the value of $U$ and plot several densities of states at $U=3.14t$ and $U=2.9t$ for rotationally-invariant and ``density-density'' models respectively on Fig. \ref{DOS2_T}. These points are chosen close to $U_c$ so that at the temperature $T=0.1t$ both models show no quasiparticle peak at Fermi level. With a decrease of temperature to $0.035t$ the three-peak structure is observed for the rotationally-invariant model while the gap persists in the ``density-density'' case. The fact that the central peak emerges with decreasing of
the temperature reveals its truly Kondo nature. This also indicates that absence of this peak is related to the lack of Kondo-type resonances in the ``density-density'' model.

\begin{figure}[t]
\begin{flushleft}
\includegraphics[width=0.96\columnwidth]{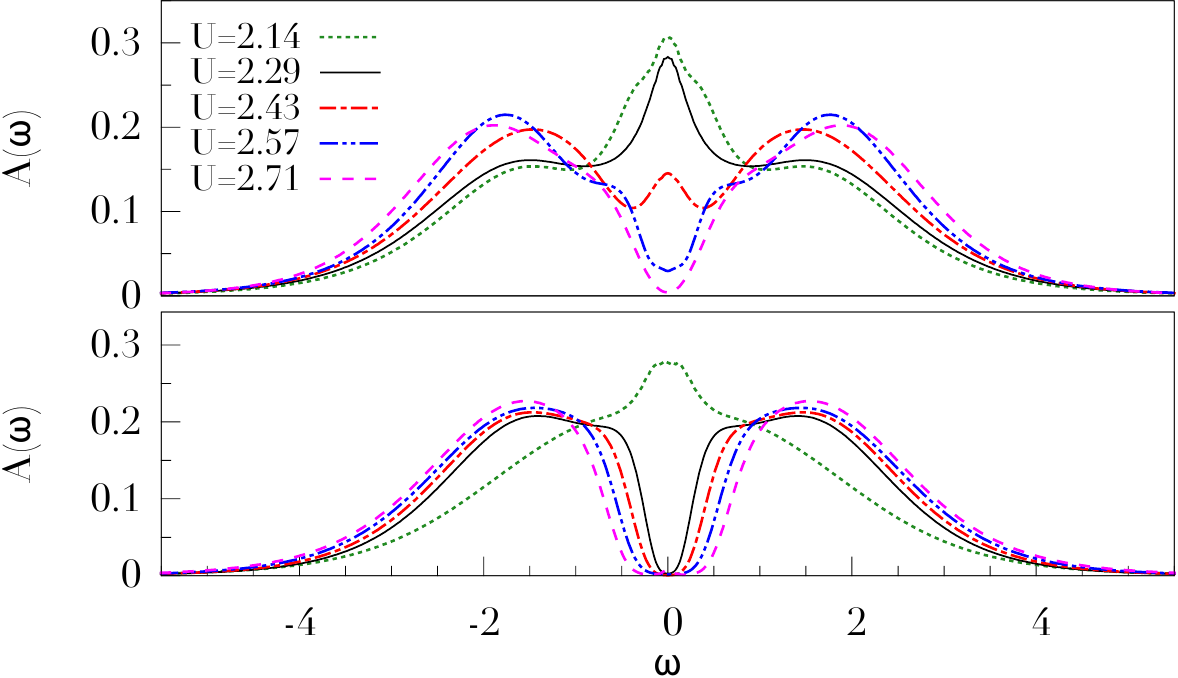} \\
\end{flushleft}
\caption{ Family of the density of states of the three-orbital Hubbard model at half-filling for various values of $U$ in the vicinity of the metal-insulator transition, $J=U/4, T=0.05t$. The curves are plotted in the same fashion as on Fig. \ref{MultipleDOS2}, upper part corresponds to rotationally invariant interaction, lower - to its ``density-density'' part.  } \label{MultipleDOS3}
\end{figure}

We conclude this part by comparing densities of states of the half-filled three-orbital model for the rotationally invariant and ``density-density'' interactions at various $U$ in the vicinity of a Mott transition on Fig. \ref{MultipleDOS3}. The curves are almost identical to the two-orbital ones and the same behavior of the central peak is seen. The main difference is that the value of critical $U$ is even more suppressed to $\approx 2.6t$ and $\approx 2.2t$ in the rotationally invariant and ``density-density'' models respectively. This reduction of critical $U$ with the number of orbitals directly follows from the formula (\ref{Umatrix}). Keeping $J=0$ restores the $U_c=4t$ value of the critical interaction strength\cite{PruschkeBulla:2005}.

A more detailed discussion on the mechanism of formation of the central quasiparticle peak is given in Appendix \ref{DF}.

\section{Discussion}
\label{discussion}

\begin{figure}[t]
\includegraphics[width=0.96\columnwidth]{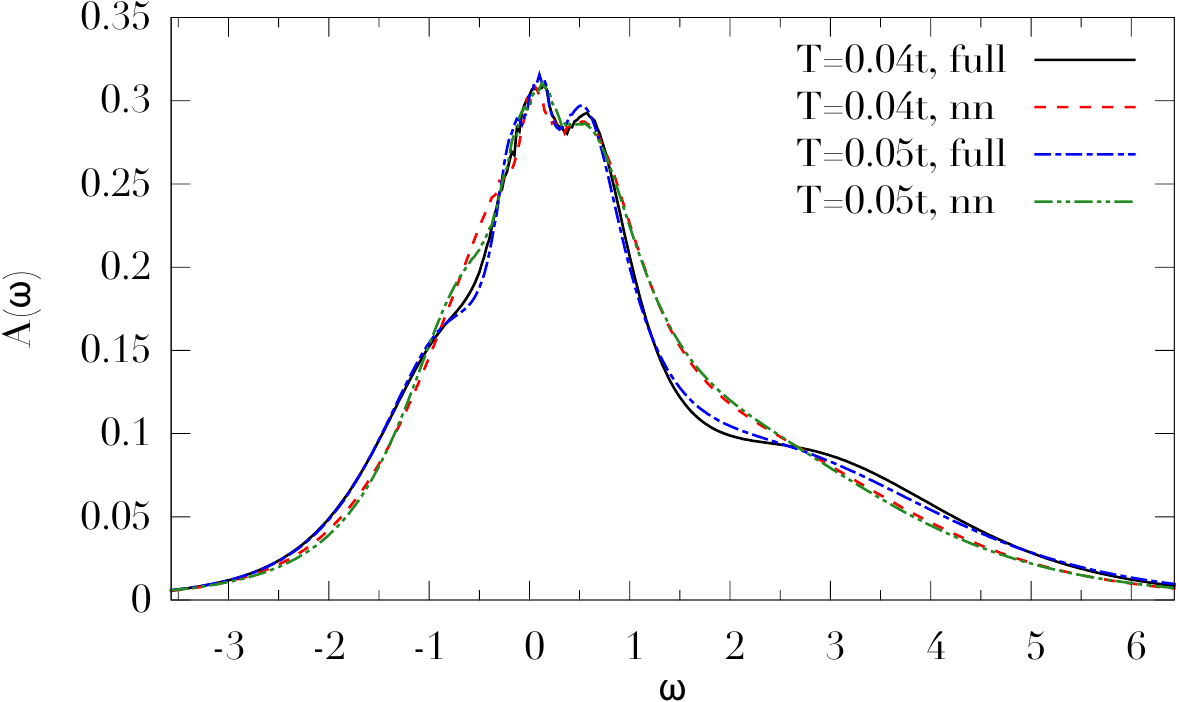}
\caption{Densities of states of the one-third filled ($2$ electrons) three-orbital model at $U=2.6t, J=U/4$ for $T=0.04$ and  $0.05t$ obtained by maximum entropy analytic continuation \cite{Sandvik:1998}.  }
\label{DOS3N2}
\end{figure}

The results obtained demonstrate a drastic difference between the two different regimes, namely for the half-filled and the one-third filled case. 
The physics of a particle-hole symmetric system is clearly of the Mott type. Hubbard bands are formed near the point of metal-insulator transition (see Figs \ref{MultipleDOS2}, \ref{DOS2_T}, \ref{MultipleDOS3}), which coincides with the divergence in the ``antiferromagnetic'' susceptibility.  On the other hand, the DOS for the one-third filling (Fig. \ref{DOS3N2}) does not demonstrate any signatures of the Hubbard band formation and the system is very far from Mott transition. We can attribute this effect to a large reduction of the effective Coulomb interaction strength from $U_{eff}=U+2J$ in the three electron case to $U_{eff}=U-3J$ for two electrons in the three orbital model \cite{MediciMravlje:2011}. 

\begin{figure}[t]
\includegraphics[width=0.96\columnwidth]{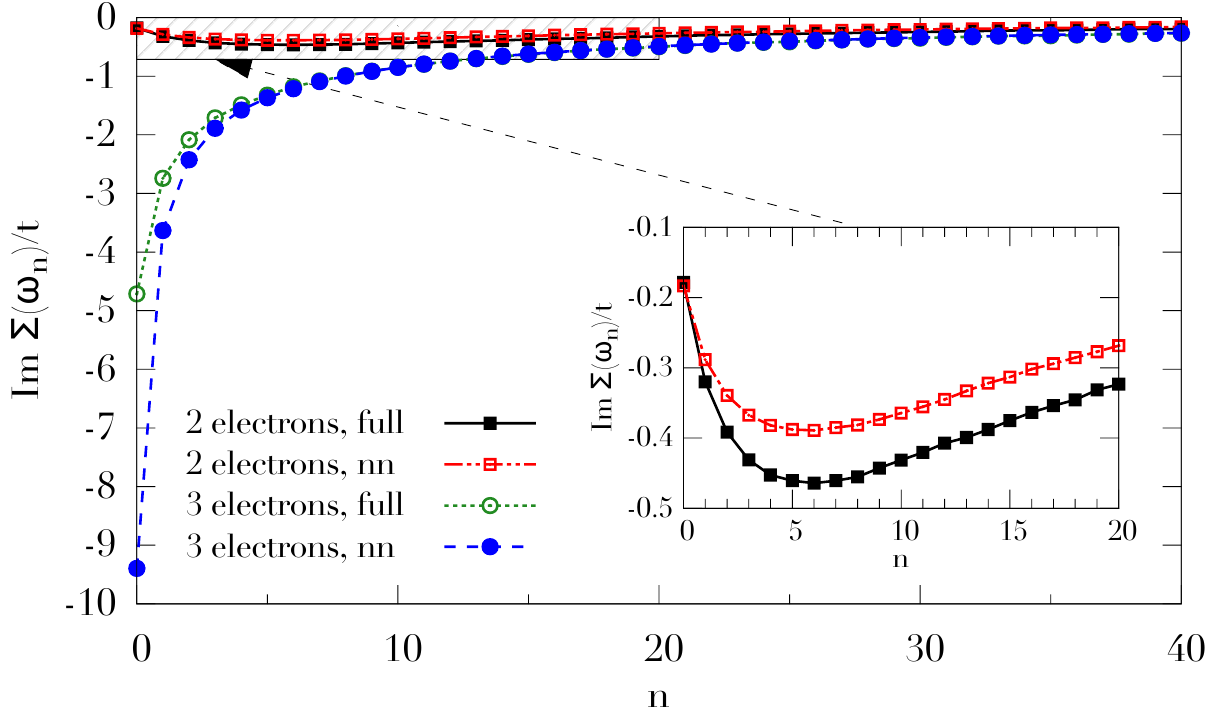} \\
\caption{Imaginary part of the self-energy in the Matsubara domain plotted as a function of the number of Matsubara frequency in the three-orbital model (\ref{RotInvU}) with one-half and one-third fillings at $U=2.6t, J=U/4, T=0.05t$. The half-filled curve shows an insulating diverging behavior, whereas the curve for a one-third filling reaches a finite value with a proximity to zero. The inset shows the curves for one-third filling zoomed at first Matsubara frequencies. }
\label{Sigma3}
\end{figure}

\begin{figure}[t]
\includegraphics[width=0.96\columnwidth]{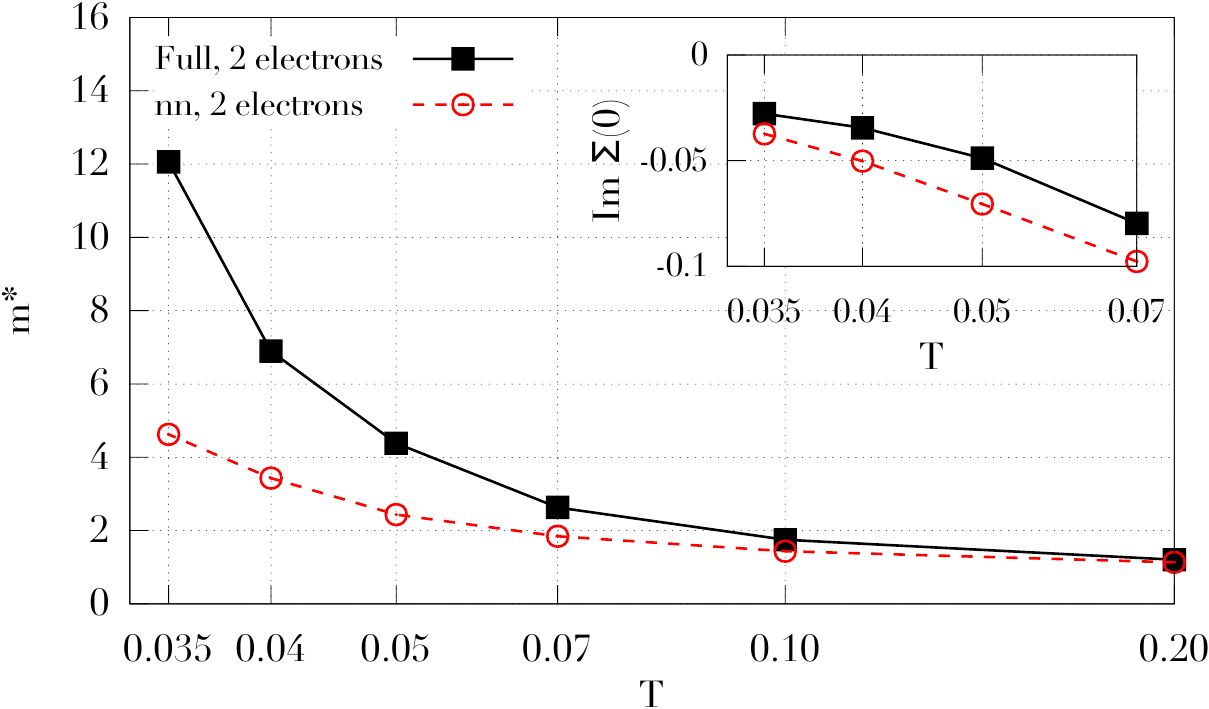}
\caption{Effective mass as a function of temperature in the one-third filled  three-orbital model (\ref{RotInvU}) at $U=2.6t, J=U/4$ for the density-density and rotationally invariant Coulomb interactions. Inset: Imaginary part of the self-energy at $\omega=0$ as a function of temperature for the same parameters. }
\label{Mass}
\end{figure}

However, it does not mean that the one-third filled system is not correlated. As the main measure of the strength of correlations, one can consider the local self-energy function $\Sigma_\omega$. Figure \ref{Sigma3} depicts ${\rm Im} \Sigma$ in Matsubara domain. For the half filling, the behavior is typical for Mott materials: $\Sigma$ tends to diverge near Fermi level. 
For one-third filling, the system is metallic with a big damping due to a strong scattering on spin fluctuations. Larger absolute value of the self-energy at the Fermi level, obtained from the extrapolation from the first Matsubara frequencies (inset of Fig. \ref{Mass}) indicates less coherent quasiparticles in the ``density-density'' model while both curves with a decrease of temperature approach zero as it is expected for the Fermi liquid and ``frozen moment'' phases \cite{WernerGull:2008, LiebschIshida:2012}.

The main curve on Fig. \ref{Mass} shows that while the introduction of the full rotationally-invariant interaction brings a minor difference to the density of states (on Fig. \ref{DOS3N2}) and the absolute value of the self-energy (in the inset) the effective mass is much more susceptible to the particular form of the interaction. This is a manifestation of the strong correlations in the system. For the density-density type of model for $J=U/4$ with the decrease in temperature the effective mass is $\approx 2.5$ times smaller than for the model with the rotationally-invariant interaction. Several mechanisms of this effect are possible. First, the stronger enhancement of an effective mass in the model with the rotationally invariant interaction can be attributed to the $1.5$ times larger degree of degeneracy of the ground state of the atomic model \cite{MediciMravlje:2011}. Beside that, in the described models correlated electrons interact with effective spin fluctuations. The Ising-like interaction produces fluctuations only for the $z$-component of a spin, whereas the Heisenberg type of interaction imposes quantum fluctuations for all spin components. In the simplified description, where the interaction between electrons is replaced with the interaction with the fluctuating classical field, one expects that the electronic self-energy behaves as the square of the field fluctuations. The latter in general are bigger for the Heisenberg than for the Ising type of fluctuations.

In the standard single-band Hubbard model, $s=1/2$ single-particle local momenta appears because of the $U$-term in the Hamiltonian. 
The peculiarity of multiorbital physics is a formation of a many-electron multiplet picture. Taking into account the exchange term $J$ with full rotational invariance is of a principal importance here, because it is responsible for the Hund's rule validity for many-electron configurations. 
The difference between the two cases can be attributed to the mechanism of local moment formation. For the $U$ term, the formation of local momenta is accompanied with the localization of electrons; Hubbard bands appear due to the charge fluctuations. On the other hand, Hund's mechanism of a local moment formation does not require 
localization; momenta can appear in a well-developed metallic state. One can refer to the terminology of `Hund's metal' introduced previously \cite{YinHaule:2011, YinHaule:2011a}.

Clearly, the difference between the rotationally invariant and Ising-type interaction is more pronounced for the Hund's metals.
Indeed, charge fluctuations almost do not contribute to the electronic Green's function, the $U$ term is not of importance, and correlation effects are related with spin fluctuations. In this situation, a particular form 
of the exchange part of Hamiltonian must affect the result strongly.
Larger effect of the symmetry of Hamiltonian on the transition temperature for the one-third filled system (Fig. \ref{cwFM}) and bigger enhancement of effective mass (Fig. \ref{Mass}) confirm this statement.

\section{Conclusion}

We conclude that the magnetism of degenerate orbital Hubbard-like models away from the half filling is dominated by the Hund's rule effects. In this case it is of crucial importance to keep the proper spin-rotational symmetry of interaction Hamiltonian for the estimation of the critical magnetic temperatures and effective masses. The corresponding single-particle properties are normally metallic while effective Coulomb interaction is small. In the half-filled case, this situation is opposite: the dominating physics is normally related with the large effective Coulomb interactions and so the antiferromagnetic transition temperature is not very sensitive to the particular symmetry of the Hund's rule term. The single particle spectrum in this case can be quite sensitive to the spin-flip terms due to the Kondo-like effects.

\begin{acknowledgments}
Authors are grateful to S. Biermann, M. Katsnelson, A. Poteryaev, S. Kirchner and E. Gull for fruitful discussions. 
The calculations were performed using the computational resources of MCC NRC ``Kurchatov Institute'' (http://computing.kiae.ru/). 
This work was supported by RFBR grants No 11-02-01443,  10-02-00046a, 10-02-00546 and 12-02-91371-CT\_a; DFG through FOR 1346. ISK acknowledges support from French ANR under projects CORRELMAT and SURMOTT. VIA acknowledges the support from the fund of the President of the Russian Federation for the support of scientific schools under grant NSH-6172.2012.2; the Program of the Russian Academy of Science Presidium ``Quantum microphysics of condensed matter''; the Russian Ministry of Education and Science under the grant No 12.740.11.0026. ANR acknowledges support from the Russian Ministry of Education and Science, grant No 07.514.12.4033. AEA and ANR acknowledge ``Dynasty'' foundation for the financial support. 
\end{acknowledgments}

\appendix
\section{Dual fermion perturbation theory}\label{DF}
In this appendix we provide the analytical formalism to explain the appearance of the central peak in the density of states of the half-filled model with a rotationally invariant interaction as compared to the ``density-density'' counterpart. It was previously noted by Pruschke and Bulla \cite{PruschkeBulla:2005} that the particular structure of the ground state of all spins aligned in the same direction in the ``density-density'' model does not allow for the appearance of the central peak. This description is clearly valid in the Kondo limit, however for the system in the strongly correlated regime the argumentation requires some extra justification. 
To proceed with this case, we employ the ``hybridization-like'' expansion in auxiliary fermionic variables (dual fermions) around an exact solution of the atomic problem\cite{KrivenkoRubtsov:2010}. The vertex functions of the atomic problem are used in the diagrammatic expansion. The appearance of the peak is connected to the divergence in the dual fermion self energy. It will be shown that in the model with a rotationally invariant interaction this divergence is guaranteed by the structure of the vertex function.

One important circumstance is that for a degenerate atomic ground state such a theory cannot be constructed straightforwardly, since the degeneracy, in general, breaks the perturbation series. 
In a work by Hafermann \textit{et al.}\cite{HafermannJung:2009} the dual fermion perturbation theory is written for a symmetric single-band Anderson model (see figure 4 of [\onlinecite{HafermannJung:2009}]). 
It has been shown that the Kondo-like resonance is correctly
reproduced within this approach only if the reference system (``isolated atom'')
is in fact not just a single impurity atom, but a cluster including the impurity and at
least one bath site (``superperturbation theory''). 
The superperturbation theory does not suffer from the degeneracy issue, because the ground state of such a cluster is a singlet.

An alternative way proposed in [\onlinecite{KrivenkoRubtsov:2010}] 
is to deal with a single isolated atom, and to break the symmetry of its ground state with a small external field, so that the perturbation series is constructed around a polarized atomic solution (a similar approach was 
proposed by Logan \cite{Logan:1998}). Such a symmetry-broken version of the theory makes a qualitative analysis feasible.

For simplicity, we consider the paramagnetic phase only (that is, put $h=0$ in  (\ref{SIAMcw})) and write the formulae for the zero temperature. 
The impurity action $S^{\text{imp}}$ is split into a sum of an atomic action corresponding to the Hamiltonian (\ref{Umatrix}) and a hybridization term (a summation over orbital and spin indices $\alpha$ is supposed): 
\begin{multline}\label{broken}
    S^{\text{imp}} = \left(S^{\text{at}} - \int\limits_{-\infty}^{+\infty}dt\ h_\alpha n_{t\alpha} \right)-\\-
        \iint\limits_{-\infty}^{+\infty} dt\ dt'\ c^*_{t\alpha}\left(\Delta_\alpha(t-t') - h_\alpha \delta(t-t') \right)c_{t'\alpha},
\end{multline}
where $\Delta_\alpha(t-t')$ is the self-consistent hybridization function. 
The integrals are taken over the axes of real times $t$.
An infinitesimal field $h_\alpha$ is introduced to break the symmetry of the atomic problem;
the result will be averaged over all possible orientations of $h_\alpha$.

Following the procedure of the dual transformation \cite{KrivenkoRubtsov:2010}, a new set of fermionic (dual) variables $f^*_{\e\alpha}, f_{\e\alpha}$ is then introduced through a Hubbard-Stratonovich decoupling of the last term in (\ref{broken}): $c^* (\Delta-h) c \to c f^* + c^* f - f^* (\Delta-h)^{-1} f$.
An average value, represented by a path integral over $c^*,c$ with a kernel $\exp(iS^\text{imp})$ now can be
rewritten in the introduced mixed representation as an integral over both $c^*,c$  and new variables $f^*,f$.
Moreover, it is possible to switch to the pure dual representation by integrating out only $c^*,c$
variables. This leads to a dual action with, in general, nonlinearities of all even orders:
\begin{equation}\label{Sd}
    S^\text{dual} = -\sum_{\alpha}\int\limits_{-\infty}^{+\infty}d\e\
        (\tilde G^{0}_\alpha(\e))^{-1} f^*_{\e\alpha}f_{\e\alpha} + V^d[f^*,f]
\end{equation}
\begin{equation}\label{Gd0}
    \tilde G^{0}_\alpha(\e) = \frac{1}{g_\alpha(\e)^{-1} - \Delta_\alpha(\e)} - g_\alpha(\e)
\end{equation}
Functions $g_\alpha(\e)$ are one-particle Green's functions of the
atomic problem: $g_\alpha(t-t') = -i\melem{\gs}{\mathbb{T}c_{t\alpha} c^+_{t'\alpha}}{\gs}$, where $\mathbb{T}$ is the time-ordering operator and $g_\alpha(\e)$ is the Fourier transform of the $g_\alpha(t-t')$,  $g_\alpha(\varepsilon$) is $\alpha$-diagonal for the models under consideration. 
Coefficients of quartic and higher terms of dual potential $V^d[f^*,f]$ are proportional to
irreducible vertex parts of the atomic problem (without fermionic legs).

It can be shown that a partition function and any observable of the physical electrons is rigorously related to the dual counterparts. In particular, for the impurity Green's function we have

\begin{multline}\label{GandSdual}
    G_\alpha(\e)^{-1}= {g_\alpha(\e)^{-1} -\Delta_\alpha(\e) - \left(g_\alpha(\e) + \tilde \Sigma^{-1}_\alpha(\e)\right)^{-1}}
\end{multline}

The dual self-energy $\tilde \Sigma_\alpha(\e)$ is defined through a standard Dyson equation for the dual
particles. Just like in the case of physical electrons, the low-frequency behavior of the self-energy for the dual fermions reflects the occurrence of an insulating gap.

We can see from (\ref{GandSdual}) that $\tilde \Sigma(0)\to\infty$ leads to a metal DOS obeying the Friedel sum rule,
and $\tilde \Sigma_\alpha(0)\to 0$ leads to the Hubbard-I solution with an insulating gap (which may be of zero width).
This fact is somewhat opposite with respect to the analogous statement for the physical
self-energy. But it comes as no surprise, since the physical $\Sigma_\alpha(\e)$ is an essential
building block in the weak coupling theory and, as it mentioned above, we are dealing with the strong coupling expansion.

The dual fermion representation allows to construct a conventional diagrammatic perturbation theory
for $\tilde \Sigma$ using $\tilde G^{0}_\alpha(\e)$ as bare lines on the diagrams, and the irreducible vertex parts as
frequency-dependent expansion parameters. All standard statements, including the Wick's theorem, the
linked-cluster theorem and the Dyson equation hold for the dual perturbation theory.
 
One can understand the dual perturbation theory for an impurity model as an alternative formulation
of the strong coupling expansion. Indeed, the zeroth order of the theory coincides with the Hubbard-I
approximation, but the dual transformation gives us Wick's theorem and therefore the ability to develop
a simple diagrammatic perturbation theory. The conventional formulation of the hybridization expansion
is lacking the Wick's theorem, because the isolated atom is an essentially nonlinear system. Thus the achievement of Wick's theorem cannot come at no price, and the price for it is the emergence of
the higher nonlinear terms in the dual action.

We are going to estimate the lowest-order contribution to the dual self-energy (``Hartree'' diagram). 
\begin{equation}\label{diagram}
    \tilde \Sigma_\alpha(\e) = \frac{i}{2\pi}\int\limits_{-\infty}^{+\infty}d\e'
        \gamma^{(4)}_{\alpha\beta\alpha\beta}(\e,\e';\e,\e') \tilde G^{0}_\beta(\e')
\end{equation}
Note that this diagram equals zero for the dual series built on top of a DMFT solution. 
However, here we construct an expansion on top of the atomic problem, and (\ref{diagram}) does not vanish.
Moreover, we will discuss in what case it diverges at the Fermi level. In accordance with the discussion above, 
such a divergence indicates a pinning of the metallic DOS.

Let us analyze expressions for the two-particle Green's function $\chi^{(4)}_{1234}$ and the corresponding vertex function $\Gamma^{(4)}_{1234}$. For real
frequencies at $T=0$ there exists a spectral representation for $\chi^{(4)}_{1234}$ in analogy
to expressions given in Refs. \onlinecite{Nozieres:1964} and \onlinecite{HafermannJung:2009}:

\begin{widetext}
\begin{equation}
    \chi^{(4)}_{1234}(\epsilon_1,\epsilon_2;\epsilon_3,\epsilon_4) =
        \sum_{\Pi}\sgn(\Pi)\sum_{klm}
            \langle \gs|\hat O_{\Pi_1}|k\rangle \langle k|\hat O_{\Pi_2}|l\rangle
            \langle l|\hat O_{\Pi_3}|m\rangle \langle m|\hat O_{\Pi_4}|\gs\rangle 
            \ \phi_{klm}(z_{\Pi_1},z_{\Pi_2},z_{\Pi_3},z_{\Pi_4})
\end{equation}
\begin{equation}
\phi_{klm}(z_1,z_2,z_3,z_4) = -i
    \frac{1}{z_1 + E_{\gs} - E_k + i\delta}
    \frac{1}{z_1 + z_2 + E_{\gs} - E_l + i\delta}
    \frac{1}{z_1 + z_2 + z_3 + E_{\gs} - E_m + i\delta},
\end{equation}
\end{widetext}
where $|\gs\rangle$ is the ground state, indices $k,l,m$ denote the eigenstates of the Hamiltonian (\ref{Umatrix}) and $\Pi$ denotes the permutation of the operators, which happens due to the time-ordering. The sum over $\Pi$ is thus a sum over all 24 permutations of operators $\hat O = \{c_1, c_2, c^+_3, c^+_4\}$.
The whole frequencies dependence of the vertex is contained in the spectral kernel $\phi_{klm}$. 
Arguments $z$ are defined as $\{\e_1,\e_2,-\e_3,-\e_4\}$.

The symmetry breaking of the atomic problem by an infinitesimal external field makes 
the ground state $\gs$ unique and so no summation over $\gs$ is performed. The $k,m$ states differ from $\gs$ in the number of particles, and 
have therefore a different energy (see Tables \ref{2bandTable}, \ref{3bandTable}). Contrary, $E_l$ can coincide or be very close to $E_{\gs}$
 (if $l$ coincides with $\gs$ or another ground state of the unperturbed atom, respectively).
This ``singular'' case is particularly interesting to us, because a resonant kernel
$(z_1+z_2+i\delta)^{-1}$ may turn into $(\e-\e'\pm i\delta)^{-1}$ in the formula (\ref{diagram})
and result in a logarithmic divergence at $\e\to0$ as soon as $\tilde G^{0}_\beta(\e')$ goes to a finite value
at the Fermi level.

The multiplet structure of the atomic ground state is of crucial importance here. First of all, if the ground state of the atomic problem is unique, a singular term in the two particle 
Green's function will inevitably correspond to $\ket{l}=\ket{\gs}$. But such singular terms cancel with the one-particle reducible part and thus do not
enter the vertex. From this we conclude that singular terms in
$\Gamma^{(4)}_{1234}$ appear, in general, for atoms with a degenerate ground state. 
Furthermore, not all of the states of the multiplet contribute to the singular part, since $\ket{l}$ is connected to $\ket{\gs}$ by 
a pair of creation-annihilation operators $\hat O_{\Pi_1}, \hat O_{\Pi_2}$. It means that a contributing term is produced only when
total spin projections of states $\ket{l}$ and $\ket{\gs}$ differ by $\pm1$.

A resonant term  $(\e-\e'\pm i\delta)^{-1}$ in the integrand of (\ref{diagram}) results in the logarithmic 
divergence of $\tilde \Sigma(\epsilon=0)$ and, consequently, in the pinning of DOS at Fermi, 
when  $\tilde G^{0}_\beta(0)$ is finite. The latter is true if both $\Delta_\beta(0)$ and $g_\beta(0)$ are finite.
When the first part of the condition
means the bare density of states being conductive, the last one is satisfied when the particle-hole
symmetry is broken in the corresponding $\beta$-mode. But as we consider solutions with broken
symmetries, there must be at least one combination of indices $\beta$ and $\ket{\gs}$ which gives
$g_\beta(0)\neq0$.

Now it is easy to see that an impurity model with the density-density interaction does not
produce a pinned solution, whereas rotationally-invariant does.
 The ground states of these models are two-fold degenerate
(see tables I and II). Components of the doublets have the same orbital moment projections but their
spin moment projections differ by $\Delta S_z=1$ for the 2-band model and by $\Delta S_z=3/2$ for
the 3-band model. There is no such a pair of operators $c_\alpha c^+_\beta$ which could connect
the components of the doublets, because the differences $\Delta S_z$ are too big. As a consequence, singular
terms in $\Gamma^{(4)}_{1234}$ are not produced by density-density interactions in the electron-hole
symmetric case and the resulting impurity DOS turns to be insulating within the chosen approximation
for the dual self-energy.

\bibliography{Cwbib2011}
\end{document}

%% file: tables.tex
\begingroup
\squeezetable
\begin{table}[ht]
    \begin{tabular*}{\columnwidth}{c|c|c|c|c} \hline \hline
    \multicolumn{5}{l}{Rotationally invariant Coulomb interaction, $N=2$} \\

        $n$   & $E$   & $S$ & $g$   & \textrm{ Eigenfunctions} \\
        \hline
        2 & $\mathbf{-2\mu+U-3J}$ & 1 & 3 & $\biket{\up}{\up}, \frac{1}{\sqrt{2}}(\biket{\up}{\down}+\biket{\down}{\up}), \biket{\down}{\down}$ \\
        
        2 & $-2\mu+U-J$ & 0 & 2 & $\frac{1}{\sqrt{2}}(\biket{\up}{\down}-\biket{\down}{\up}), \frac{1}{\sqrt{2}}(\biket{\updown}{0}-\biket{0}{\updown})$ \\
        2 & $-2\mu+U+J$ & 0 &  1 & $\frac{1}{\sqrt{2}}(\biket{\updown}{0}+\biket{0}{\updown})$ \\ \hline \hline

\multicolumn{5}{l}{``Density-density'' Coulomb interaction, $N=2$} \\

        $N$   & $E$   & $|S_z|$ & $g$  & \textrm{Eigenfunctions} \\
\hline 
        2 & $\mathbf{-2\mu+U-3J}$ & 1 & 2 &  $\biket{\up}{\up}, \biket{\down}{\down}$ \\  
        2 & $-2\mu+U-2J$ & 0 & 2 & $\biket{\up}{\down}, \biket{\down}{\up}$ \\
        2 & $-2\mu+U$ & 0 & 2 & $\biket{\updown}{0}, \biket{0}{\updown}$\\
\hline
        
\end{tabular*}
    \caption{Eigenfunctions of the rotationally invariant (top) and ``density-density'' (bottom) Hamiltonians, which belong to the subspace of the double electronic occupancy for the two-orbital model. $n$ denotes number of electrons, $E$ --- energy of the eigenstate, $S$ --- total spin, $Sz$ - spin projection on the z-axis, $g$ --- degree of degeneracy. The ground state is typeset in bold.}
\label{2bandTable}
\end{table}
\begin{table}[ht]
\begin{longtable}{c|c|c|c|c} \hline\hline
\multicolumn{5}{l}{Rotationally invariant Coulomb interaction, $N=3$} \\
        $n$   & $E$   & $S$ & g & \textrm{Sample eigenfunctions} \\
        \hline
           2     & $-2\mu+U-J$     & 0     & 5 &
            $
                \begin{matrix}
                    \frac{1}{\sqrt{2}}(\triket{\up}{\down}{0}-\triket{\down}{\up}{0})\text{ + 2 st.} \\
                    \frac{1}{\sqrt{2}}(\triket{\updown}{0}{0}-\triket{0}{0}{\updown})\text{ + 1 st.}
                \end{matrix}
            $\\ \hline
        2     & $\mathbf{-2\mu+U-3J}$    & 1     & 9 &
            $
                \begin{matrix}
                    \triket{0}{\up}{\up} \text{ + 5 st.} \\
                    \frac{1}{\sqrt{2}}(\triket{\up}{\down}{0}+\triket{\down}{\up}{0}) \text{ + 2 st.}
                \end{matrix}
            $\\ \hline
        2     & $-2\mu+U+2J$    & 0     & 1 &
            $\frac{1}{\sqrt{3}}(\triket{\updown}{0}{0}+\triket{0}{\updown}{0}+\triket{0}{0}{\updown})$\\ \hline
        3     & $-3\mu+3U-6J$   & 1/2   & 10 &
            $
                \begin{matrix}
                    \frac{1}{\sqrt{2}}(\triket{\up}{\up}{\down}-\triket{\down}{\up}{\up}) \text{ + 3 st.}\\
                    \frac{1}{\sqrt{2}}(\triket{\updown}{0}{\up}-\triket{0}{\updown}{\up}) \text{ + 5 st.}
                \end{matrix}
            $\\ \hline
        3     & $-3\mu+3U-4J$   & 1/2   & 6 &
            $
                    \frac{1}{\sqrt{2}}(\triket{\updown}{0}{\up}+\triket{0}{\updown}{\up})\text{ + 5 st.}
            $\\ \hline
        3     & $\mathbf{-3\mu+3U-9J}$   & 3/2   & 4 &
            $
                \begin{matrix}
                    \triket{\up}{\up}{\up}\text{ + 1 st.}\\
                    \frac{1}{\sqrt{3}}(\triket{\up}{\up}{\down}+\triket{\up}{\down}{\up}+\triket{\down}{\up}{\up}) \\ \text{ + 1 st.}
                \end{matrix}
            $\\ \hline\hline
\multicolumn{5}{l}{``Density-density'' Coulomb interaction, $N=3$} \\
        $N$   & $E$   & $|S_z|$ & $g$  & \textrm{Sample eigenfunctions} \\
        \hline
        2     & $-2\mu+U$ & $0$ & 3 &
            $\triket{\updown}{0}{0}$ + 2 st.\\
        2     & $-2\mu+U-2J$ & $0$ & 6 &
            $\triket{\up}{\down}{0}$ + 5 st.\\
        2     & $\mathbf{-2\mu+U-3J}$ & $1$ & 6 &
            $\triket{\up}{\up}{0}$ + 5 st.\\
        3     & $-3\mu+3U-5J$ & $1/2$ & 12 &
            $
               \triket{\updown}{0}{\up}
            $ + 11 st.\\
        3     & $-3\mu+3U-7J$ & $1/2$ & 6 &
            $\triket{\up}{\up}{\down}
            $ + 5 st.\\
        3     & $\mathbf{-3\mu+3U-9J}$ & $3/2$ & 2 &
            $\triket{\up}{\up}{\up}$ + 1 st.\\\hline
        \end{longtable}\addtocounter{table}{-1}
        \caption{Eigenfunctions of the rotationally invariant (top) and ``density-density'' (bottom) Hamiltonians, which belong to the subspaces of two- and three- electrons occupancy for the three-orbital model (\ref{RotInvU}). The notations are the same as in Table \ref{2bandTable}, apart that only the eigenstates nonequivalent under spin and orbital space rotations are included in the table (the number of the other equivalent         states is denoted with ``st''). The ground states for each electronic occupancy are typeset in bold.}
\label{3bandTable}
\end{table}
\endgroup